\documentclass{article}
\usepackage{graphicx}
\usepackage{amsthm,amssymb,amsmath}
\usepackage[T1]{fontenc}
\usepackage[margin=1in]{geometry}
\usepackage[noblocks]{authblk}
\usepackage[margin=1cm]{caption}
\renewcommand{\figurename}{Fig.}

\title{Two-photon interference between disparate sources for quantum networking}
\date{}
\author[1,2]{A. R. McMillan\thanks{email: alex.mcmillan@bristol.ac.uk}}
\author[3]{L. Labont\'e\thanks{email: laurent.labonte@unice.fr}}
\author[1,4]{A. S. Clark}
\author[1]{B. Bell}
\author[3]{O.~Alibart}
\author[3]{A. Martin\thanks{Currently with the GAP (University of Geneva)}}
\author[2]{W. J. Wadsworth}
\author[3]{S. Tanzilli}
\author[1]{J. G. Rarity}

\affil[1]{{\it \small Centre for Communications Research, Department of Electrical and Electronic Engineering, University of Bristol, Merchant Venturers Building, Woodland Road, Bristol, BS8 1UB, UK}}
\affil[2]{{\it \small Centre for Photonics and Photonic Materials, Department of Physics, University of
Bath, Claverton Down, Bath, BA2 7AY, UK}}
\affil[3]{{\it \small Laboratoire de Physique de la Mati\`ere Condens\'ee, CNRS UMR 7336,
Universit\'e de Nice -- Sophia Antipolis, Parc Valrose, 06108 Nice Cedex 2, France}}
\affil[4]{{\it \small Centre for Ultrahigh Bandwidth Devices for Optical Systems (CUDOS), Institute of Photonics and Optical Science (IPOS),
School of Physics, University of Sydney, NSW 2006, Australia}}

\begin{document}
\maketitle

\noindent {\bf Quantum networks involve entanglement sharing between multiple users. Ideally, any two users would be able to connect regardless of the type of photon source they employ, provided they fulfill the requirements for two-photon interference. From a theoretical perspective, photons coming from different origins can interfere with a perfect visibility, provided they are made indistinguishable in all degrees of freedom. Previous experimental demonstrations of such a scenario have been limited to photon wavelengths below 900\,nm, unsuitable for long distance communication, and suffered from low interference visibility. We report two-photon interference using two disparate heralded single photon sources, which involve different nonlinear effects, operating in the telecom wavelength range. The measured visibility of the two-photon interference is $80 \pm 4\%$, which paves the way to hybrid universal quantum networks.} 

\vspace{-20 pt}

\section*{}

\begin{figure}[h!]
\begin{center}
\resizebox{0.5\columnwidth}{!}{%
\includegraphics{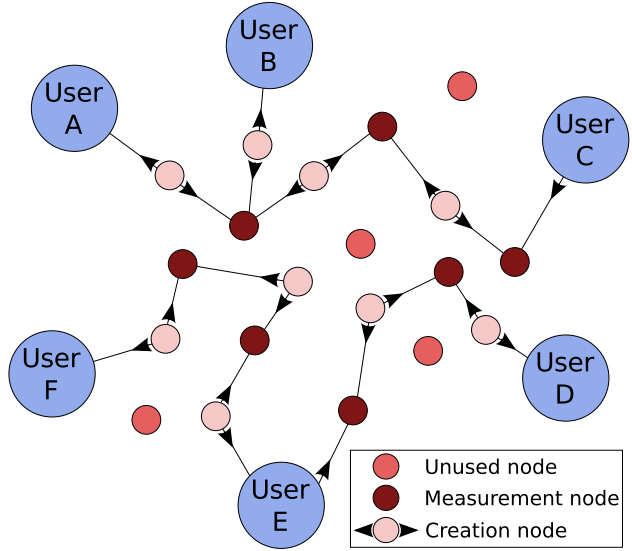}}
\caption{Schematic of a quantum network connecting distant users through the connection of various nodes where entanglement is created or measured. The arrows define the direction to which entanglement is distributed.\vspace{-0.8cm}}
\label{fig_QN}
\end{center}
\end{figure}

\noindent Connecting distant quantum devices enables quantum networking applications, such as distributed quantum computing~\cite{Grover_telecomp} and quantum key distribution (QKD)~\cite{Gisin_QKD,Scarani_QKD_2009}. Quantum networks (QNs) mostly exploit entanglement as a primary resource distributed between various partners that are not necessarily connected by a direct link. As sketched in \figurename{~\ref{fig_QN}}, some nodes are capable of emitting and receiving (entangled) quantum bits of information (qubits), while others are dedicated to creating or measuring entanglement. Combining different pairs of entangled qubits by a local quantum operation, as in quantum relays~\cite{Martin_12} and repeaters~\cite{Q_repeaters_sangouard}, allows qubits that never physically meet to become entangled. This operation is called ``entanglement swapping''~\cite{DeRiedmatten_swapping_05} which allows, by chaining such operations, the creation of end-to-end quantum links between arbitrarily-spaced users. From the general perspective, the specific choice of the entanglement carriers and associated observables does not matter~\cite{Weihs_Photonic_ent_01}. Atoms and ions are preferred for qubit storage and manipulation at specific locations~\cite{Lvovsky_09,Q_repeaters_sangouard}, while photons are ideal carriers for transferring qubits over relatively long distances, as well as for on-chip manipulation~\cite{Tanzilli_genesis_2012, MartinPuce}. Telecom wavelength photons can travel along low-loss optical fibres, with filtering and routing advantageously handled using high performance fibre components. Recent progress has identified Fourier-transform limited picosecond photons as ideal candidates for realistic QN applications~\cite{Abo2010}.

From the experimental side, entanglement swapping relies on two-photon interference, a purely quantum effect which occurs when two identical photons enter opposite input ports of a beam-splitter (BS). This effect, also known as coalescence, causes the two photons to exit through the same output port of the BS and therefore leads to an absence of measured coincidences, or so-called Hong-Ou-Mandel (HOM) dip, between detectors placed at the outputs of the device~\cite{HOM1987,Legero}. Here, the type of generation process does not matter, since only the produced single photon properties have to be considered for optimizing the interference visibility. Up to now, this effect has been extensively studied using various types of sources, based either on nonlinear crystals~\cite{Abo2010,Clark2011,Kaltenbaek_Inter_Indep_06}, quantum dots~\cite{Santori,patel,Flagg}, NV centers in diamond~\cite{Harald}, single atoms~\cite{Beugnon}, atomic ensembles~\cite{Felinto}, or trapped ions~\cite{Maunz}. However, typically the two photons originate either from different sources with the same generation process, or from the same source.  While it is unclear at this stage which, if any of the currently developed photonic source technologies will be proven to be the most suitable for 
future QNs, it may be the case that the ideal solution will encompass multiple types of sources.

We report in this article a two-photon interference experiment involving two disparate photon-pair sources based on different nonlinear generation processes, a major step toward entanglement sharing over a QN. It is very recently that the first hybrid experiment has shown a measured interference visibility of 16\% between two photons produced by a quantum dot and a nonlinear-crystal-based photon pair source~\cite{Polyakov2011}. This proof-of-principle experiment was limited by strong mismatches between the single photon properties, and was at an inappropriate wavelength for quantum networking. Here, we exploit photon-pair sources based on three-wave mixing in a periodically poled lithium niobate waveguide (PPLN/W) and four-wave mixing in a microstructured fiber (MF). These types of sources have already shown their capabilities in terms of pair generation efficiency and entanglement quality~\cite{Tanzilli_PPLW_02,Rarity:05,Fulconis_nonclassical_07}. In the following, we describe our experimental set-up, in which photons from these two sources are made to interfere. We demonstrate a two-photon interference visibility of $80 \pm 4\%$ between telecom wavelength photons from these sources and discuss the relevance of this result in view of QN applications.

\section*{Results}

\subsection*{Design and characterisation of the photon sources}

\noindent Both sources have been engineered to provide paired photons emitted near 1550\,nm and 810\,nm. The near-visible wavelength is particularly suitable for local operation since it is compatible with high efficiency detectors and quantum logic gate operations~\cite{Tanzilli_genesis_2012,OBrien07122007,ClarkCNOT}. On the other hand, the telecom wavelength corresponds to the point of minimum loss in telecommunication fibre and allows standard high performance fibre components to be utilised~\cite{Abo2010}.

\begin{figure}[h]
\begin{center}
\resizebox{0.7\columnwidth}{!}{%
\includegraphics{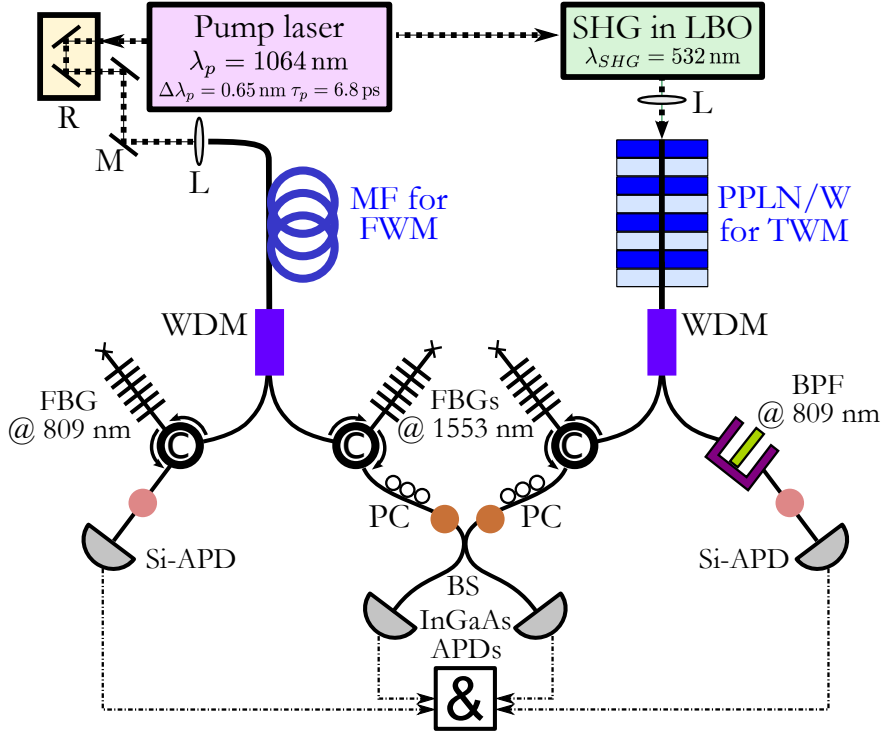}}
\caption{Setup combining the MF-based and PPLN/W source (left and right side of diagram). The 1553\,nm photons from both sides are filtered using FBG filters and are combined at a 50:50 fused fibre beam-splitter (BS). The two sources act as heralded single photon sources thanks to the detection of the 809\,nm photons which herald the idler photons at 1553\,nm.
R: retroreflector; M: mirror; WDM: wavelength division multiplexer; C: circulator; FBG: fibre Bragg grating filters; PC: polarization controller; APD: avalanche photodiode; \&: FPGA logic system for recording four-fold coincidences.}
\label{Set-up}
\end{center}
\end{figure}

As shown in \figurename{~\ref{Set-up}}, two separate output beams from a picosecond pulsed, 1064\,nm fibre laser are respectively directed towards a PPLN/W and a section of MF. Both devices are single-mode at telecom wavelengths, enabling the phase-matching of a single nonlinear process in each system and thereby avoiding any additional background noise within the bandwidth of interest. The source shown on the left-hand side of \figurename{~\ref{Set-up}} consists of a 20\,cm section of MF designed to exploit the $\chi^{(3)}$ nonlinearity of silica and obtain four-wave mixing (FWM), leading to paired photons near 810/1550\,nm from two photons at 1064\,nm \cite{mc-oe-17-6156}. Its length has been optimized to minimize the temporal walk-off between the pump pulse and the generated pairs of photons to obtain the best efficiency-to-bandwidth ratio. On the right-hand side, a 2\,cm-long PPLN/W fabricated through the soft-proton exchange technique~\cite{Tanzilli_PPLW_02} permits photon pair production by three-wave mixing (TWM) in the $\chi^{(2)}$ nonlinear medium, at the same signal and idler wavelengths as above, upon the annihilation of a single photon at 532\,nm. In order to satisfy the requirements of energy matching for the TWM process, the initial 1064\,nm beam is frequency doubled using a bulk lithium triborate (LBO) crystal in a temperature stabilised oven at $150^\circ \rm{C}$, before being launched into the PPLN/W.

For each source, we separate pairs of signal and idler photons into single-mode fibres using a standard wavelength demultiplexer (WDM), before applying spectral filtering in order to select energy-matched pairs of photons at wavelengths of 809.2\,nm and 1553.3\,nm respectively. In the case of the idler photons, which are used to demonstrate HOM interference, narrowband spectral filtering is implemented using a combination of a low loss circulator and fibre Bragg grating (FBG). These gratings are designed to reflect photons lying within a 600\,pm bandwidth, satisfying the requirements of the Fourier transform criterion for spectrally pure photons, defined by the pulse duration of the pump laser.

\begin{table}[!ht]
\caption{Detailed operating parameters for both the MF and the PPLN/W sources.\label{Tab.1}}
\vskip 0.25cm
\begin{center}
\begin{tabular}{l||c|c}
\textbf{}&\textbf{MF}&\textbf{PPLN/W}\\
\hline
\hline
Sample length (cm) & 20 & 2\\
\hline
Idler wavelength (nm)&1553.3&1553.3\\
\hline
Signal wavelength (nm)&809.2&809.2\\
\hline
Idler FBG bandwidth (nm)&0.6&0.6\\
\hline
Signal filtering bandwidth (nm)&0.15&0.5\\
\hline
Trigger counts (kHz)&100&80\\
\hline
Coincidence counts (kHz)&1&0.5\\
\hline
Coincidence-to-accidental ratio &$\sim20$&$\sim20$\\
\hline
Pump power (mW)&250&2\\
\hline
\end{tabular}
\end{center}
\end{table}

Both sources were characterized independently in an operating regime of a maximum of 0.05 created pairs per laser pulse. \tablename{~\ref{Tab.1}} summarizes the characteristics of these two sources. The PPLN/W exhibits a very high efficiency and consequently a pump power of 2\,mW (at 532\,nm) is enough to reach the required operating regime. Furthermore, on the PPLN/W side, the emission process does not induce Raman noise, which is responsible for additional background counts in the MF source. The fibre source is inherently better adapted for coupling to standard single-mode fibres than the PPLN/W, which leads to higher coincidence count rates.

\subsection*{Theoretical interference visibility}

\noindent The visibility ($V$) of the HOM dip resulting from the interference of idler photons from the two sources can be written as~\cite{Rbook,Zukowski}:
\begin{equation}\label{visibility}
V=\frac{1}{\sqrt{1+\frac{\Delta t ^{2}_{PPLN/W}}{2 \Delta \tau ^{2}}+\frac{\Delta t ^{2}_{MF}}{2 \Delta \tau ^{2}}}},
\end{equation}
where $\Delta \tau $ is related to the coherence time defined by the FBG bandwidth, while  $\Delta t_{PPLN/W}$ and $\Delta t_{MF}$ are the effective duration of the photon wave-packet for the PPLN/W and the MF source respectively. Note that $\Delta t$ depends on the laser pump operation regime as described by Aboussouan et al.~\cite{Abo2010}, and corresponds either to the detection timing jitter for the continuous wave regime, or to the idler pulse duration for the pulsed regime. Furthermore, during the propagation along the 2\,cm-long PPLN/W, the group velocity dispersion between signal (809\,nm) and pump photons induces a 6\,ps broadening of the idler photon wavepacket due to the walk-off effect. 
Taking into account this additional uncertainty on wavepacket arrival time,  a theoretical interference visibility, given by Eq.~\ref{visibility}, of 83\% is expected.

\subsection*{Demonstration of two-photon interference} 

\noindent To maximize the interference visibility, the 1553\,nm single photons have to be rendered indistinguishable in all their degrees of freedom. In our case, the polarization modes are made identical by using fibre polarization controllers (PC) placed just before the BS. In addition, maximal spatial mode overlap at the BS is ensured by the use of a single-mode fused fibre coupler. In order to observe HOM interference, the two photons are required to enter the BS simultaneously within their coherence time (given by the bandwidth of the idler arm FBGs). The relative arrival time of the photons at the BS is adjusted using a retroreflector~(R) mounted on a motorized translation stage, which is placed in the path of the laser pulses in front of the MF photon pair source. The signal and idler photons from both sources are detected using avalanche photodiode detectors (APD). Four-fold counts in which all four APDs detect a photon simultaneously are monitored and recorded using coincidence counting electronics (\&).   

\begin{figure}[h]
\begin{center}
\resizebox{0.8\columnwidth}{!}{\includegraphics{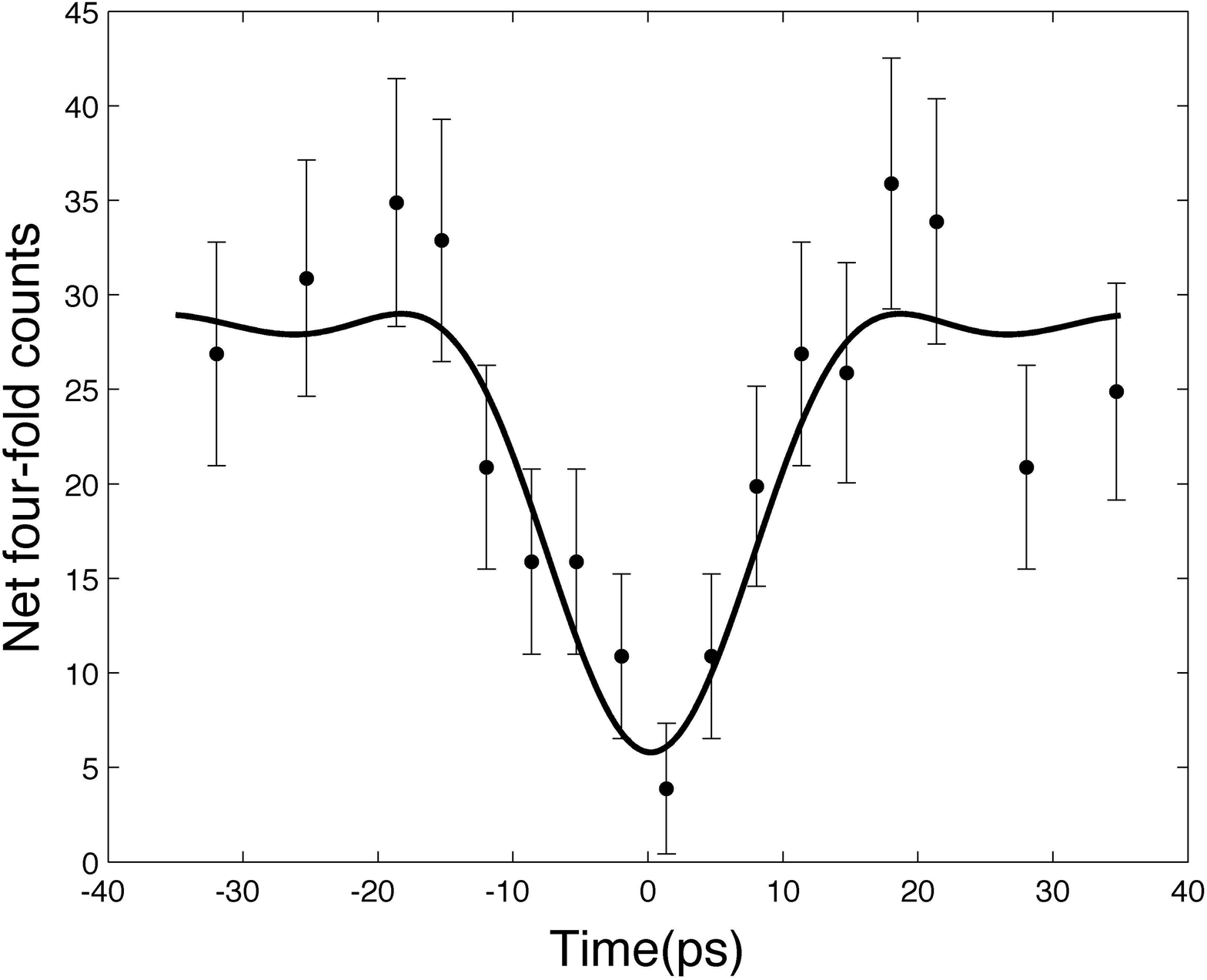}}
\caption{Net four-fold coincidence rate as a function of the relative delay between idler photons at the BS, adjusted using the retroreflector on the MF source side. The acquisition time for data was 56 minutes at each point. The error bars are calculated based on a Poissonian distribution for the measured counts. The line shows a fit to the data, which is a sinc-squared function due to the near square shape of the spectral filtering provided by the narrowband FBGs.}
\label{Results}
\end{center}
\end{figure}

When the two idler photons are made indistinguishable, a 70\% reduction in the raw four-fold coincidence rate, the HOM dip~\cite{HOM1987}, is demonstrated. The net four-fold coincidence rate is shown in \figurename{~\ref{Results}}, and has been corrected by subtracting the accidental background four-fold coincidence events produced by each source individually. The background contribution for each source was determined by blocking the idler arm path of the other source and counting the measured four-fold coincidence events over a period of 3.5 hours. The combined background count rate from the two sources was found to be 0.145 counts per minute, implying a total background count of 8.12 for each measured data point. The net visibility of the four-fold coincidences is $80 \pm 4\%$, well within agreement with the expected visibility ($83\%$).

Moreover, the measured temporal width of the HOM dip of $16.5 \pm 2$\,ps agrees well with the expected value of 17\,ps. We could implement narrower FBGs in order to increase $\tau_{c}$, and consequently, the visibility. Reducing the bandwidth of the idler FBGs down to 200\,pm would not have a significant impact on the counting rate, while it would increase the visibility up to 97\%.

\section*{Discussion}

\noindent In this paper, we have implemented a set-up used to observe non-classical HOM interference from two different types of source, an all-fibre  source and a source based on a $\chi^{(2)}$ crystal. The operation of these sources relies on two different nonlinear processes (FWM and TWM). An $80 \pm 4\%$ net interference visibility was achieved. We have identified the main factor limiting this visibility, and a reasonable way to overcome this restriction. We stress that this work will allow any pair of sources to be used together in QN applications, provided the photons generated are indistinguishable in all degrees of freedom. In this paper, the choice of the nature of the source is motivated by the need to provide proof of the feasibility of a scalable quantum network. More generally, it is the first step towards realising applications in future QNs encompassing multiple types of photon sources, such as quantum relays based on entanglement swapping operations \cite{Halder07}.\\

\section*{Methods}

\subsection*{Pump source and spectral filtering}

\noindent The pump laser used in this experiment was a custom dual output model, in which a single oscillator was used to provide synchronised output pulses from two separate laser amplifier stages (Fianium Ltd. \textit{FemtoPower FP-1060-0.25}). Both outputs from the system supplied 7\,ps-duration pulses at the wavelength of $\lambda_{p}=$~1064\,nm, with a spectral bandwidth of $\Delta \lambda_{p}=$ 0.7\,nm and a repetition rate of 80\,MHz. Based on the properties of the pump laser, 600\,pm bandwidth FBGs were selected for filtering the idler photons in both sources, corresponding to a coherence time of 13.4\,ps for the interfering photons. The narrowband idler FBGs were tuned using strain to shift the central wavelength of reflection, both in order to optimally match the applied spectral filtering to the generated idler wavelengths for FWM and TWM, and to ensure that the filtering profiles of the two sources were identical. For the non-interfering signal photons in the two sources, the requirements for spectral filtering were less stringent. In the MF source, a 150\,pm bandwidth FBG was used, so that the filtering bandwidths were energy matched for both the heralding and interfering photon to help minimise the influence of uncorrelated background noise. In the PPLN/W source, where the generated noise in the idler channel was lower, the signal photons were filtered using a slightly wider 500\,pm free-space interference filter, in order to minimise the optical loss and maintain reasonable counting rates. 

\subsection*{Detector arrangement for coincidence measurements}

At the output of the set-up, four-fold photon coincidences between signal and idler photons from both sources were recorded using two sets of silicon (Si) and indium gallium arsenide (InGaAs) APD units (Perkin Elmer \textit{SPCM-AQR-14} and ID Quantique \textit{id201}, respectively), all connected to a field programmable gate array (FPGA) coincidence logic system. Each InGaAs detector was advantageously gated using the heralding signal from a Si-APD after detection of an 809\,nm photon. This allows the two systems to be operated as heralded single photon sources, and therefore limits the overall noise in the four-fold coincidence rate.

\vspace{-20 pt}

\renewcommand\refname{}

\section*{Acknowledgements}
\noindent The authors thank Florian Kaiser for fruitful discussions. Financial support from the Royal Society, the CNRS, the European ICT-2009.8.0 FET project QUANTIP (grant 244026), the ERC project QUOWSS (grant 247462), the EU project Q-Essence (grant 248095), the University of Bristol, the Australian Research Council Centre of Excellence (CUDOS, project number CE110001018), and the University of Nice - Sophia Antipolis, is acknowledged.

\section*{Contributions}
\noindent S.T. and J.G.R. conceived the idea for this project. The PPLN photon source was developed by L.L., O.A., A.M. and S.T., and A.R.M., W.J.W. and J.G.R. were responsible for the development of the MF source. A.R.M., L.L., A.S.C. and B.B. conducted the interference experiment. The paper was prepared by L.L., O.A. and S.T. and reviewed by all authors. 

\section*{Competing financial interests}
\noindent The authors declare no competing financial interests.

\end{document}